\documentclass[aps,prl,reprint,superscriptaddress]{revtex4-2}
\usepackage{graphicx}
\usepackage{amsmath}
\usepackage{amssymb}
\usepackage{mathrsfs}
\usepackage[T1]{fontenc}
\usepackage{calligra}
\usepackage{array}
\usepackage[dvipsnames]{xcolor}
\usepackage{float}
\usepackage{soul}
\usepackage{bm}
\usepackage[toc,page]{appendix}
\usepackage{braket}
\usepackage{wrapfig}
\usepackage[colorlinks=true, linkcolor=blue, urlcolor=blue,citecolor=blue]{hyperref}

\global\long\def\av#1{\left\langle #1\right\rangle }%

\global\long\def\ket#1{\left|#1\right\rangle }%

\global\long\def\bra#1{\left\langle #1\right|}%

\global\long\def\b#1{\mathbf{#1}}%

\global\long\def\cal#1{\mathcal{#1}}%

\global\long\def\an#1#2{\left\{  #1,#2\right\}  }%

\usepackage{printlen}
\uselengthunit{cm} 

\hypersetup{colorlinks,linkcolor=blue,citecolor=blue,urlcolor=blue}

\begin{document}

    \title{Towards Defect Engineering of Orbital Magnetoelectric Effects}
    
    \author{João M. Alendouro Pinho}
    \thanks{These authors contributed equally to this work.}
    \affiliation{Centro de Física das Universidades do Minho e Porto Departamento de Física e Astronomia,
Faculdade de Ciências, Universidade do Porto, 4169-007 Porto, Portugal}
    \author{Henrique P. Veiga}
    \thanks{These authors contributed equally to this work.}
    \affiliation{Centro de Física das Universidades do Minho e Porto Departamento de Física e Astronomia,
Faculdade de Ciências, Universidade do Porto, 4169-007 Porto, Portugal}
    \affiliation{School of Physics, Engineering and Technology and York Centre for Quantum Technologies, University of York, YO10 5DD, York, United Kingdom}
    \author{Alessandro Veneri}
    \affiliation{Dipartimento di Ingegneria dell’Informazione, Elettronica e Telecomunicazioni, Università La Sapienza, Via Eudossiana, 18, 00184 Roma, Italy}
    \affiliation{School of Physics, Engineering and Technology and York Centre for Quantum Technologies, University of York, YO10 5DD, York, United Kingdom}
    \author{Daniel Passos}
    \affiliation{Centro de Física de Materiales, Universidad del País Vasco, 20018 San Sebastián, Spain}
    \author{Ivo Souza}
    \affiliation{Centro de Física de Materiales, Universidad del País Vasco, 20018 San Sebastián, Spain}
    \affiliation{Ikerbasque Foundation, 48013 Bilbao, Spain}
    \author{Bruno Amorim}
    \affiliation{Centro de Física das Universidades do Minho e Porto Departamento de Física e Astronomia,
Faculdade de Ciências, Universidade do Porto, 4169-007 Porto, Portugal}
    \author{João M. Viana Parente Lopes}
    \affiliation{Centro de Física das Universidades do Minho e Porto Departamento de Física e Astronomia,
Faculdade de Ciências, Universidade do Porto, 4169-007 Porto, Portugal}
    \author{Aires Ferreira}
    \email{aires.ferreira@york.ac.uk}
    \affiliation{School of Physics, Engineering and Technology and York Centre for Quantum Technologies, University of York, YO10 5DD, York, United Kingdom}

\begin{abstract}

Orbital magnetization is a fundamental measure of time-reversal symmetry breaking in solids that is accessible by experiments. However, unlike the well-studied equilibrium case, the understanding of orbital magnetism resulting from external stimuli such as light and electric fields is still in its infancy. Here we employ a novel real-space framework to investigate charge-to-orbital conversion in a metallic orbital antiferromagnet beyond the conventional Bloch-band picture. We find that common point defects give rise to strongly enhanced time-odd magnetoelectric effects and, depending on their local symmetry, further activate otherwise-forbidden time-even responses. Moreover, we introduce a local marker that provides spatially resolved information on the orbital magnetization carried by non-equilibrium states. Real-space maps reveal that point defects create nanoscale hotspots of orbital magnetization leading to a profound modification of the bulk magnetoelectric response, including a sign reversal of the net orbital moment. Our findings may open up the possibility of engineering linear-response orbital moments with atomic-scale precision by point defects and impurities.
    
\end{abstract}
    
\maketitle

The orbital degree of freedom of electrons is attracting growing interest driven by fundamental advances, such as the discovery of orbital magnetism in moiré flat-band systems \cite{Lu_2019_orbital_mag, Polshyn_2020_ChernI, Tschirhart_2021_orb_mag, Persky_2026_orbital_mag} and rare earths \cite{Hao_26_rare_earths}, as well as its potential for fast control of magnetic states with promising applications in data storage and computing technologies  \cite{Lee_OrbitalTorque_2021, Lee_OrbitalXC_2026}. 

On a microscopic level, orbital magnetism emerges from the interplay of interactions, spin-orbit coupling and crystal symmetry, playing an essential role in diverse phenomena such as magnetic anisotropy \cite{Bruno_89_TB_MCA,Gambardella_03,Donati_13, Ovesen_25},   magnetic circular dichroism \cite{Thole_92_XMCD}, and valley-polarized topological phases of matter in twisted bilayer graphene \cite{Liu_2019_TBG, Bultinck_20_TBG, Pierce_21, Yu_22}. The orbital magnetic moment of periodic crystals enjoys an elegant description in terms of Berry-curvature-like quantities, which incorporates itinerant orbital contributions absent in the atom-centred approximation \cite{Chang_96_ModernT, Thonhauser_05_ModernT, Xiao_2005_ModernT, Xiao_2010_RMP}. Combined with momentum-space data obtained from electronic structure calculations \cite{Lopez_12_ModernT}, this so-called ``modern theory'' has provided a practical and accurate framework to compute the macroscopic orbital magnetization of realistic systems \cite{Lopez_12_ModernT,Hanke_16,Nikolaev_14_ModernT,Vidarte_26}.
Recently, this approach has revealed important insights into interaction-driven orbital ferromagnetism  in correlated quantum  phases realized in twisted bilayer transition-metal dichalcogenides \cite{Liu_26_ModernT} and in rhombohedral graphene superlattices \cite{Han_23_pentalayer,sheekey_26}.

Whilst the research focus has been concentrated on equilibrium orbital magnetization, the orbital physics of non-equilibrium states may be equally significant from both fundamental \cite{Bernevig_05} and applied \cite{Lee_21_orb_torque,Ding_22,Choi_23} perspectives.  For example, it is predicted that current-induced orbital moment accumulation---the orbital analog of the inverse spin galvanic effect \cite{Ivchenko_1978,Ivchenko_1989,Aronov_1989,Edelstein_1990,Si_05,Benitez_20}---occurs in systems with momentum-space orbital textures \cite{Park_11,Kim_2012,Yoda_18,zhongGyrotropicMagneticEffect2016}, allowing their use to control magnetization dynamics \cite{Xi_18,Nikolaev_24}. 

An intriguing phenomenon occurs in antiferromagnetic metals with parity-time ($\mathcal{PT}$) symmetry \cite{bourges_novel_2011,he_berry_2012,gao_orbital_2018}, whose equilibrium Fermi surfaces lack an orbital texture, but are believed to display a non-equilibrium orbital moment entirely driven by quantum geometry \cite{Xiao_21,Ocana_23,shinadaQuantumTheoryIntrinsic2023,Cullen_2026}.  However, the role of realistic impurities and defects in both the equilibrium and non-equilibrium orbital magnetization remains largely unexplored. Understanding how extrinsic orbital transport mechanisms \cite{Liu_24,Veneri_25}, disorder-induced quantum phase transitions \cite{Li_09,Avedis_25} and sharp scattering resonances \cite{Veneri_25,Vidarte_26} may modify the clean-system paradigm of magnetoelectric effects \cite{Xiao_21,Ocana_23,shinadaQuantumTheoryIntrinsic2023,Cullen_2026} is crucial for advancing our knowledge of non-equilibrium orbital magnetism and has the potential to uncover new applications. 

\begin{figure*}[!t]
\centering \includegraphics[width=1\textwidth]{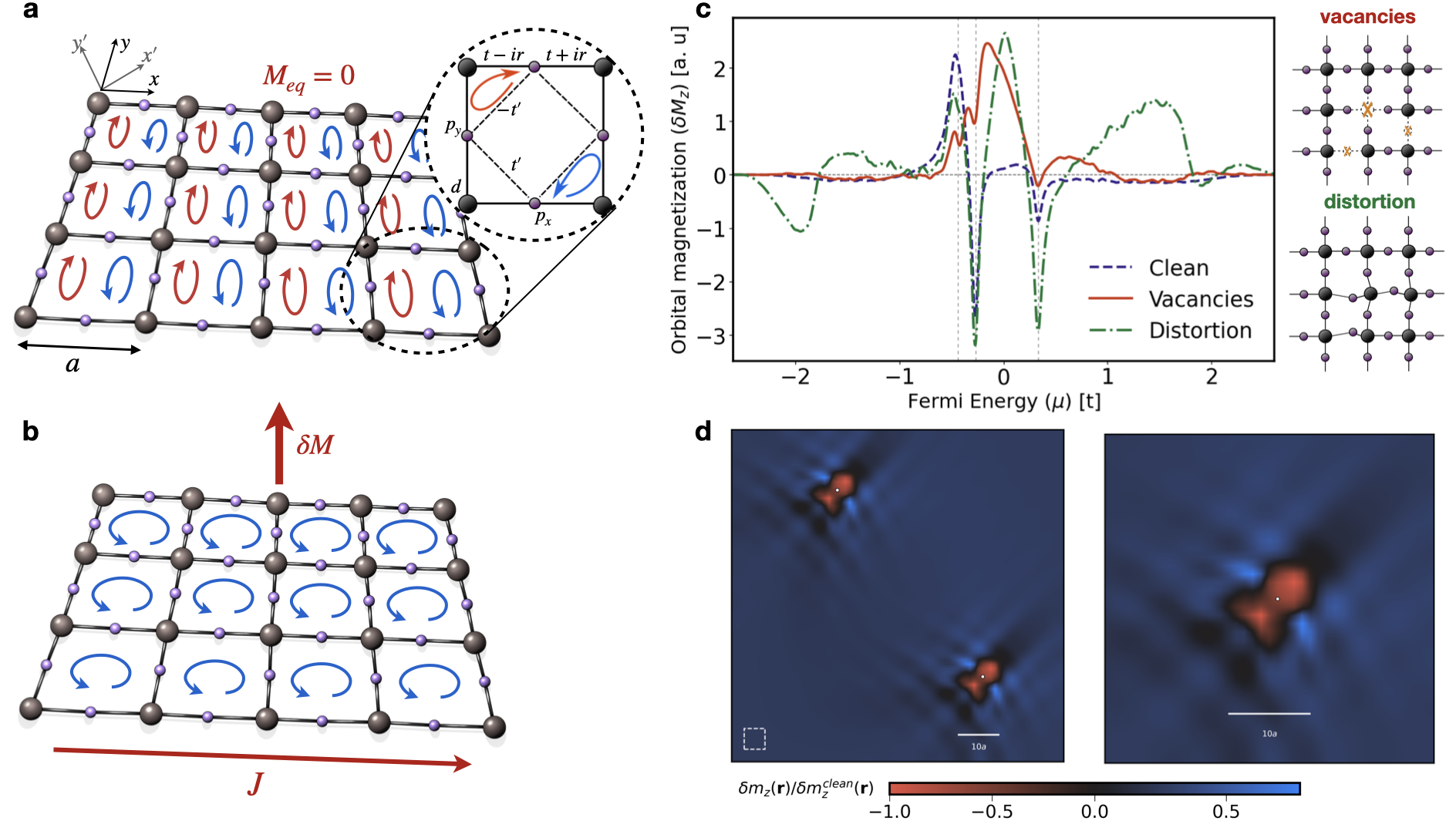} \caption{\textbf{Non-equilibrium magnetization $\delta \textbf{M}$ in broken translation-invariance
systems}. \textbf{a} Equilibrium antiferromagnetic loop-current order
in the $\mathrm{CuO_{2}}$ plane. Circulating currents generate
local orbital moments of alternating sign, resulting in zero
net orbital magnetization. The inset shows the unit cell of the model
  . \textbf{b} Under dc driving, the loop-current texture acquires a ferromagnetic component. \textbf{c} Current-induced orbital magnetization $\delta M_z$ for external field applied along the  $y\prime$-axis. The clean response (dashed line) exhibits sharp features near
  the Dirac points. Introducing a small concentration of vacancies, $N_{c}=2\%$  (solid line) produces large non-perturbative effects, including sign reversal of magnetization near $\mu=-0.27t$. Dilute polar disorder also significantly modifies the non-equilibrium state (dotted-dashed line). \textbf{d} Spatial maps of the coarse-grained normalized local orbital marker $[\delta m_z(\mathbf r)\,/\,\delta m^{\mathrm{clean}}_z(\mathbf r)]$ for the Fermi energy $\mu=0.33t$, normalized by the clean-system response. The white dashed square indicates the spatial averaging window. Near a vacancy, the local marker changes sign relative to the surrounding bulk due to the emergence of nano-domains of opposite orbital magnetization.}
\label{Fig1} 
\end{figure*}

In this work, we exploit a novel real-space framework introduced by us in Ref. \cite{Pinho_PRR} to demonstrate that common types of defects can be used to engender a variety of  orbital magnetoelectric effects that are absent in pristine crystals. To illustrate this concept, we focus on a lattice model of a 2D antiferromagnetic metal with hidden loop-current order \cite{bourges_novel_2011,gao_orbital_2018,he_berry_2012} depicted in Fig. \ref{Fig1}a. This system features, in the clean limit, a strong magnetoelectric response near the Dirac points of the low-energy band structure \cite{shinadaQuantumTheoryIntrinsic2023}, making it an ideal testbed to investigate the extent to which defects and disorder can be exploited to enhance and  tailor non-equilibrium orbital magnetization. Towards this end, we will consider in this work  two types of common defects, namely, vacancies and short-range polar bond disorder; see Figs. \ref{Fig1}(b-c). The bare Hamiltonian is the same as that employed in the Bloch-space study of Shinada \emph{et al.} \cite{shinadaQuantumTheoryIntrinsic2023}, which provides a useful benchmark for the clean translationally invariant limit. It consists of Cu \(d\) orbitals and O \(p_x,p_y\) orbitals on a Lieb lattice, with complex Cu--O hopping amplitudes generating the loop current texture and oxygen--oxygen hopping governing the low-energy band structure; see Fig. \ref{Fig1}a. 

The non-equilibrium orbital magnetization is extracted from the circulating component of a spatially dispersive current  $\mathbf J(\mathbf q,\omega)=\mathbf J^t(\mathbf q,\omega)+\mathbf J^M(\mathbf q,\omega)$ using a newly developed real-space method  \cite{Pinho_PRR}. Here,  $\mathbf J^t(\mathbf q,\omega)$ and $\mathbf J^M(\mathbf q,\omega)=i\,\mathbf q\times\delta\mathbf M(\mathbf q,\omega)$ are, respectively, the Fourier transform of the transport and magnetization current densities induced by a longitudinal electric field with frequency $\omega$ and spatial modulation wavevector $\mathbf{q}$. The central quantity of this approach is   the  linear-in-wavevector conductivity tensor, $\sigma_{ab,c}(\omega)$, which describes $\mathbf J^M(\mathbf q,\omega)$ to linear order in $\mathbf  q$ and through which the bulk magnetization induced  by a dc electric field (denoted here by $\delta \mathbf M$) can be directly extracted \cite{Pinho_PRR}. The rank-3 conductivity tensor in the dc limit ($\omega\rightarrow 0$) can be conveniently cast in real space in terms of the velocity operator $\hat{\b{v}}$, the position operator $\hat{\b{r}}$ and retarded/advanced Green's functions $\hat{G}^{{\rm R/{\rm A}}}(\varepsilon)=(\varepsilon-\hat{H}\pm i0^{+})^{-1}$, with $\hat H$ a generic single-particle Hamiltonian, all expressed in the basis of localized orbitals,
\begin{equation}
\sigma_{ab,c}=\frac{ie^2}{h}\mathrm{Re}\av{\hat{\Lambda}_{ac}\partial_{\varepsilon}\hat G^{\rm R}(\varepsilon)\hat{v}_{b}-\hat{v}_{a}\partial_{\varepsilon}\hat G^{\rm R}(\varepsilon)\hat{\Lambda}_{bc}}_{f},
 \label{eq:sigma_trace}
\end{equation}
where $\hat{\Lambda}_{ab}\,=\,\an{\hat{v}_{a}}{\hat{r}_{b}}$,
\begin{equation}
\langle \hat{O}\rangle _{f}\,=\,\frac{1}{\Omega}\int\!\!d\varepsilon\,f(\varepsilon)\,\mathrm{Tr}\left[\Delta \hat G\left(\varepsilon\right)\hat{O}\right],
 \label{eq:O_average}
\end{equation}
$\Delta G\left(\varepsilon\right)\,=\,G^{\rm R}\left(\varepsilon\right)-G^{\rm A}\left(\varepsilon\right)$, $f\left(\varepsilon\right)$ is the Fermi-Dirac distribution function and $\Omega$ is the volume (or area, in 2D). For additional details and derivation, see accompanying paper \cite{Pinho_PRR}. 

\begin{figure*}[!t]
\centering \includegraphics[width=1.0\textwidth]{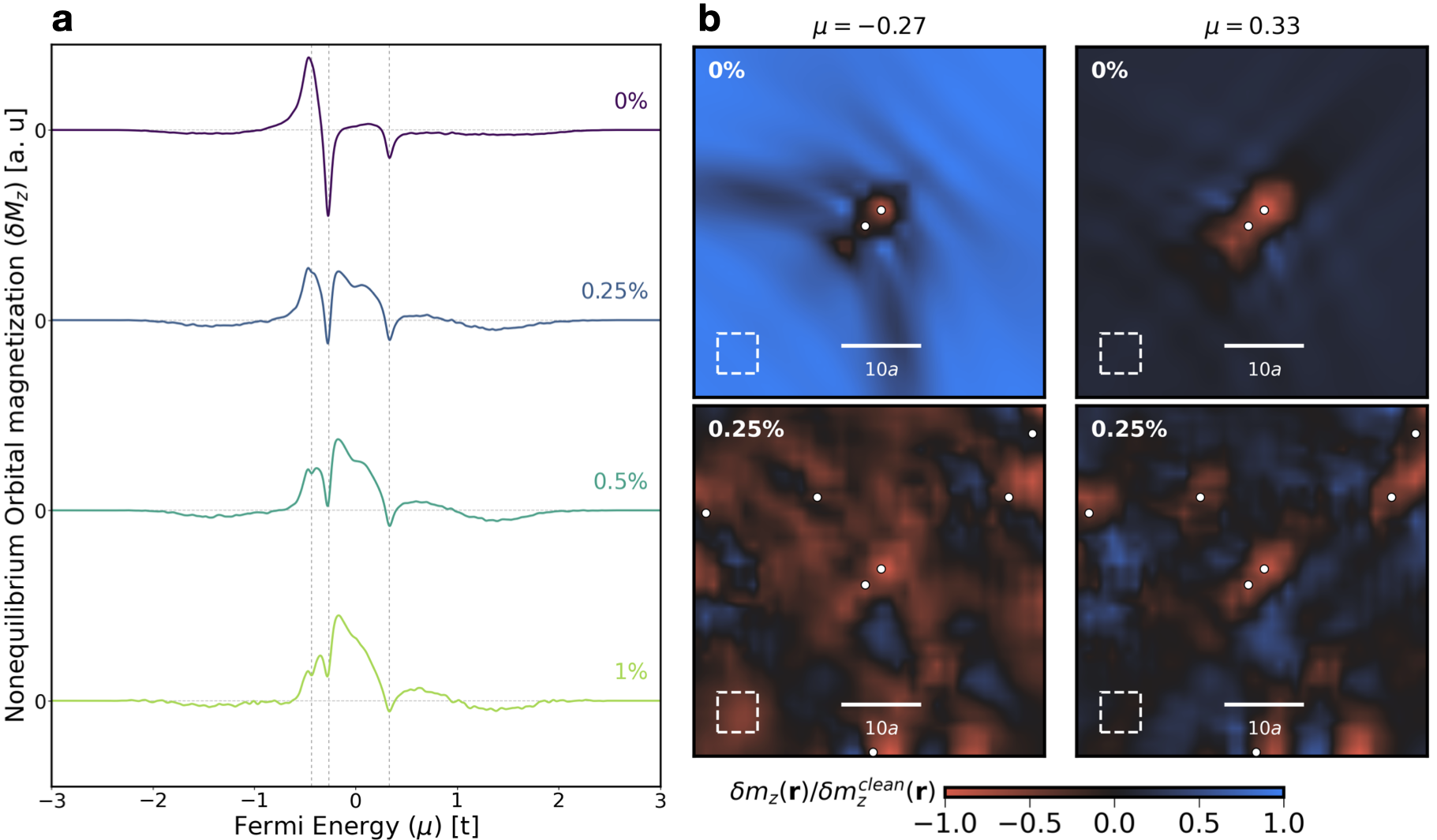} \caption{\textbf{Vacancy-induced orbital magnetization $\delta \textbf{M}$}. Non-equilibrium OM as a function of Fermi energy for increasing vacancy concentration. Curves are vertically offset for clarity, with the concentration indicated on the right. Dashed vertical lines mark the Dirac-point energies of the pristine \(\mathrm{CuO}_2\) model. Even low vacancy disorder strongly reshapes the response: for concentrations between \(0.25\%\) and \(0.5\%\), the magnetization reverses sign near \(\mu=-0.27t\), whilst a broad finite response develops in the energy window between the two higher-energy Dirac points.
\textbf{b} Spatial maps of the normalized local magnetization marker, $\delta m_z(\mathbf r)\,/\,\delta m_z^{\mathrm{clean}}(\mathbf r)$, for $\mu=-0.27t$ and $\mu=0.33t$. The upper row shows the response of two nearby vacancies in an otherwise clean system, whilst the lower row shows the same vacancy pair embedded in a disordered background with vacancy concentration $0.25\%$. White circles mark vacancy positions, dashed squares indicate the coarse-graining window, and scale bars represent $10a$. At $\mu=-0.27t$, where the macroscopic response reverses sign, the background magnetization also changes sign over an extended region. By contrast, at $\mu=0.33t$ the response remains dominated by localized hotspots of OM. This contrast indicates that vacancy-induced quantum interference is particularly strong near $\mu=-0.27t$, where it produces an extended contribution to $\delta \mathbf{M}$  rather than a simple superposition of isolated-defect responses.}.
\label{Fig2} 
\end{figure*}

To determine the impact of atomic defects on $\delta {\bf M}$ of large samples of the above model, we evaluate Eq. (\ref{eq:sigma_trace}) through large-scale Chebyshev expansions of the Green's function operators combined with stochastic trace methods (see Methods section in End Matter). Our focus is on systems with low concentration of point defects, represented either by vacancies or by local polar distortions, placed at random positions (see End Matter for details on implementation). Such forms of disorder are ubiquitous in condensed matter and are known to strongly impact the local and macroscopic properties of their host systems \cite{Rutter_07,Yazyev_07,Alloul_09,Ferreira_11,Ugeda_11,Hong_15,Jiang_18,Barja_19,Joucken_21,Wang_23}. We note that oxygen vacancies have been directly observed in cuprates \cite{hartman_direct_2019}, while local strain fields and defect-induced atomic relaxations can produce distortions of the Cu--O bond environment \cite{mundet_local_2020,oyanagi_local_2007}. More generally, both Cu and O defects are expected to occur in the stoichiometric CuO$_2$ planes of cuprates depending on the sample preparation method and conditions \cite{Richard_04,Kang_07}.

We start by calculating $\delta {\bf M}$ of a system with vacancy defects equally distributed among the Cu and O sites. We note that vacancy defects preserve on average the original $\mathcal {PT}$ symmetry of the model. The calculations are carried out for a finite lattice with open boundary conditions comprised of $2048\times 2048$ unit cells (see End Matter). The large system size allows us to simulate  low defect concentrations, whilst avoiding undesired finite-size effects. To validate our real-space methodology, we first compute the intrinsic magnetoelectric response of the pristine system. Figure \ref{Fig1}c shows that our real-space approach  accurately reproduces the results recently obtained in Ref. \cite{shinadaQuantumTheoryIntrinsic2023} using a Bloch-state formalism. This agreement is expected, as our formulation reduces to the established theory of magneto-electric effects for systems with perfect translation symmetry (see accompanying paper \cite{Pinho_PRR} for more details). Next, we go beyond the usual picture of a periodic crystal and investigate the magnetoelectric response of a disordered copper-oxide plane with 2\% concentration of vacancies. The results are averaged over more than a thousand independent random defect configurations to best capture the macroscopic behavior of $\delta \mathbf M$ (see  End Matter). The full Fermi energy dependence of the disorder-averaged $\delta {\bf M}$ at $T=0$ is illustrated in Fig. \ref{Fig1}c. These results show drastically modified magnetic properties by the disorder landscape, including a strongly enhanced magnetoelectric response in the spectral region between $\mu=-0.27t$ and $\mu=0.33 t$. 

A particularly intriguing feature is the sign reversal of the non-equilibrium orbital magnetization (OM) which occurs near the second Dirac point, located at $\mu=-0.27 t$. This shows that vacancy defects can stabilize non-equilibrium loop-current textures with opposite sign to that of the clean system, which we will  explore below. Non-equilibrium states far from charge neutrality are also impacted. This can be seen in the spectral region $0.33 t<\mu\lesssim t$, where vacancies facilitate the emergence of a significant OM. We also find that the general behavior of $\delta M_z(\mu)$ is sensitive to the orbital character of defects. In the  Supplemental Material \cite{note_SM}, we elucidate this point by addressing the case of oxygen vacancies. 

Yet more surprising features are revealed when considering other types of disorder, especially those breaking $\mathcal{PT}$ symmetry. To demonstrate this, we investigate systems with short-range polar bond disorder. For a direct comparison, we set the concentration of polar defects to 2\% (see End Matter for additional details). The current-induced OM exhibits a strikingly modified behavior as shown in Fig. \ref{Fig1}c (note, for example, the activation of a strong response for $\mu \approx -2t$ and $\mu\approx 1.5t$). Apart from these global changes, disorder can offer another way of tailoring the magnetoelectric response through defect-induced symmetry breaking. In particular, polar bond disorder is seen to activate otherwise symmetry-forbidden time-even contributions to $\delta \mathbf M$ because its statistical distribution breaks \(\mathcal{PT}\) symmetry, even after disorder averaging. We explore this in more detail in the End Matter. 

Equally strong effects are observed for relatively low  defect concentrations. This scenario is of great experimental relevance for material systems where point defects are scarce. Figure \ref{Fig2}a shows the evolution of $\delta {\bf M}$ with vacancy concentration, where new features in the magnetoelectric signal are seen to emerge already for concentrations as low as $0.25$\%. To investigate the possible mechanisms behind these results, we introduce a local marker $\delta m(\mathbf{r}_{\alpha})=k_{ba}(\mathbf{r}_{\alpha})E_{b}$,
defined by resolving in real-space the local contributions to the magnetoelectric response kernel (here, $\mathbf{r}_{\alpha}$ indicates the position of orbital $\alpha$).  Inserting a complete set of position states into Eqs. (\ref{eq:sigma_trace})-(\ref{eq:O_average}), we isolate the local contributions to the rank-3 conductivity tensor 
\begin{equation}
\begin{split}
\sigma_{ab,c}(\mathbf r_\alpha)
=&\,
\frac{i e^2}{h}\int d\varepsilon\, f(\varepsilon)\,
\mathrm{Re}\,
\bra{\mathbf r_\alpha}
\Delta \hat G(\varepsilon)
\hat\Lambda_{ac}
\partial_\varepsilon \hat G^{\rm R}(\varepsilon)
\hat v_b
\\
&\hspace{1.cm}
-
\Delta \hat G(\varepsilon)
\hat v_a
\partial_\varepsilon \hat G^{\rm R}(\varepsilon)
\hat\Lambda_{bc}
\ket{\mathbf r_\alpha}.
\end{split}
\end{equation}
By construction, our local marker $k_{ba}\left(\mathbf{r}_{\alpha}\right)$  inherits the gauge invariance of the magnetoelectric tensor \cite{note_SM}. It also exhibits the expected bulk asymptotics of a spatial response marker, in close analogy to the local Chern marker  \cite{bianco_orbital_2013,note_SM}. However, whereas the Chern marker over a finite-size domain averages to zero \cite{bianco_orbital_2013}, the sum of $k_{ba}(\mathbf{r_\alpha})$ over all lattice sites yields the macroscopic magnetoelectric coefficient and thus has a well-defined operational meaning.

We now apply the local marker to resolve the local contributions to $\delta {\bf M}$ in real space. For presentation purposes, the local marker is averaged over a small spatial window of side length $5a$, where $a$ is the lattice spacing. This suppresses sublattice-scale oscillations whilst preserving the overall structure of the magnetoelectric response \cite{note_SM}. Figure \ref{Fig1}d shows the coarse-grained local marker $m_z(\mathbf{r}_\alpha)$ for a system with only two (far-apart) Cu vacancies and $\mu=0.33t$. This real-space simulation reveals that the local magnetoelectric response has several interesting properties. First, the response is strongly anisotropic and concentrated around the defects, decaying to its bulk (pristine) value at large distances. Second, the OM signal near the vacancies is greatly enhanced and opposite-in-sign with respect to the background (pristine) value. Clearly, these defects nucleate nano-domains of strong magnetoelectric response. Such OM hot spots  have characteristic sizes extending up to 10 lattice spacings. Third, the presence of fringes emerging from the nano-domains suggests that quantum interference effects may play a role in systems with a finite concentration of defects. Note that in this example the magnetoelectric response of the two Cu vacancies does not interfere. Next, we map out the OM signal for systems with low defect concentration. Selected results are summarized in Fig. \ref{Fig2}b (bottom panels) and shown together with the corresponding spatial maps for the case of isolated vacancies (top panels). For $\mu=0.33t$, the spatial map shows nano-domains of opposite-in-sign $\delta {\bf M}$ centred around each point defect, explaining the decrease in the net OM seen in Fig. \ref{Fig3}d. Whereas the occasional isolated vacancies retain their characteristic shape, most  nano-domains are seen to hybridize, creating puddles of non-equilibrium OM. We attribute this unique effect to a coherent multiple scattering mechanism evidenced by the fast evolution of the local marker with   defect concentration. To further corroborate this observation, we compute the real-space maps for  $\mu=-0.27t$. Note that at this energy the macroscopic OM is reversed with respect to the clean-system case (see Fig. \ref{Fig1}c). Strikingly, the sign reversal of the macroscopic response is accompanied by a qualitative change in the real-space structure of the marker. The OM signal is no longer confined around the defects, but spreads through the disordered background (Fig. \ref{Fig2}b). This supports our earlier interpretation that the magnetoelectric response of these systems is governed by quantum interference effects rather than by a mere additive sum of independent vacancy contributions. Further evidence of inter-vacancy interference is provided in the Supplemental Material \cite{note_SM}.

While conventional extrinsic mechanisms of OM generation are typically formulated within an independent-scatterer picture, such a semiclassical description cannot by itself account for the numerically exact (within resolution) results reported here. Instead, the qualitative reconstruction of the real-space magneto-electric response and its correlation with the macroscopic OM sign-reversal point to coherent multiple-defect interference as an essential ingredient. Note, in particular, the strong nonperturbative evolution of the OM signal with defect concentration in Fig. \ref{Fig2}a.  Notwithstanding, we expect that single-impurity scattering mechanisms, such as orbital skew scattering \cite{Veneri_25}, to be sufficient to describe the extrinsic part of non-equilibrium OM  in systems with ultra-low defect concentrations \cite{rouKineticOrbitalMoments2017,sahin_pancharatnam-berry_2018-1,Liu_24,Veneri_25,sarkar_extrinsic_2026,cullen_theory_2026}. In such a dilute-disorder scenario, the Fermi-surface contribution to the bulk magneto-electric coefficients scales with the scattering time akin to the familiar spin Hall and inverse spin galvanic effects in spin-orbit-coupled transport (i.e. $\delta M_z \propto  n^{-1}$, with $n$ the impurity areal density). Interestingly, in the present $\mathcal{PT}$-symmetric  model of a cuprate layer, the conventional orbital skew scattering is inoperative due to the absence of a Fermi-surface orbital texture \cite{Milletari_17,Perkins2024_review}. Instead, anomalous scattering processes, such as asymmetric spin-orbit precession \cite{Huang_16}, are expected to govern the OM physics in the small $n$ limit. Further discussion on the microscopic orbital scattering mechanisms underlying OM generation in weakly disordered systems is provided in the accompanying paper \cite{Pinho_PRR}, where we explore a Haldane model with a weak random potential.

\textit{Conclusions and outlook - } Taken together, our results demonstrate that atomic-scale disorder can qualitatively reshape nonequilibrium orbital magnetism. We have established this in a $\mathcal{PT}$-symmetric antiferromagnetic CuO$_2$ monolayer, where even low concentrations of random point defects strongly modify the net orbital magnetization of current-carrying states. In particular, the macroscopic OM in samples with vacancy defects exhibits rich behavior, including a sign reversal with respect to the clean-system response for vacancy concentrations as low as 0.5\%. The importance of this finding is two-fold. First, it demonstrates that the magnetoelectric response of the pristine crystal \cite{shinadaQuantumTheoryIntrinsic2023} is insufficient to understand non-equilibrium orbital states in realistic systems, where atomic-scale defects are ubiquitous. Second, it challenges the assumption that the impact of disorder on magnetoelectric effects can be understood within the relaxation time approximation \cite{Salemi_19,Hara_20,He_20,Leiva_23,Burgos_24} or even perturbative disorder methods beyond the weak-disorder limit \cite{Levitov_85,Arfi_92,Liu_25}. Our results show that rather than acting only as a source of lifetime broadening \cite{Salemi_19,Hara_20,He_20,Leiva_23,Burgos_24}, point defects modify the orbital magnetoelectric response non-perturbatively, which gives rise to novel effects. 

The orbital response of point defects has been further elucidated by introducing a local marker that enables access to   local information. Remarkably, our studies of this local marker reveal that vacancy defects nucleate nano-scale domains of OM. For certain energy windows, most prominently near the two higher-energy Dirac cones of the cuprate model, these nano-domains feature enhanced magnetoelectric signals exceeding that of the bulk response. Furthermore, the real-space maps show that the sign reversal of the macroscopic response is accompanied by a drastic qualitative change in the vacancy-induced magneto-electric response, from predominantly localized defect OM textures to an extended pattern spanning the disordered background. These results strongly point to coherent quantum interference between individual defect-induced responses as the mechanism underlying the bulk OM sign-reversal. This possibility is experimentally relevant because point defects can be introduced and controlled via ion \cite{Ugeda_11,Jiang_18} and electron beam irradiation \cite{Robertson_23}, among other methods \cite{Wang_23}. We expect the strong local magnetoelectric responses reported here to have technological potential as tunable nano-scale sources of orbital torques, and open up pathways to engineer non-equilibrium orbital states on the atomic scale.

\section{Acknowledgements}
J.M.A.P., B.A., H.P.V. and J.M.V.P.L. acknowledge support from Fundação para a Ciência e a Tecnologia (FCT, Portugal) through
the Strategic Funding UID/04650/2025, Grant No. UID/PRR/04650/2025 and Grant No. UID/PRR2/04650/2025. Further support from Fundação para a Ciência e a Tecnologia
(FCT, Portugal) is acknowledged through the project QPI-MoireS No. 2023.17745.ICDT (B.A.,
J.M.A.P., J. M. V. P. L. and H.P.V) and Grants
No. 2023.02155.BD (J.M.A.P.) and No. 2024.00560.BD (H.V.P). H.P.V. acknowledges computational resources from the Laboratory for
Advanced Computing at the University of Coimbra. A.F.,
J.M.V.P.L., J.M.A.P. and H.P.V. acknowledge computational
resources from the University of York High-Performance Computing service,
Viking. D.P. acknowledges funding from Horizon Europe under the
Marie Skłodowska-Curie grant agreement No.101206626. I.S. acknowledges
support from Grant No.PID2021-129035NB-I00, funded by
MCIN/AEI/10.13039/501100011033 and by ERDF/EU. A.V. thanks
Roberto Raimondi for fruitful discussions.

\section{Data and Code Availability}

The datasets for all figures and the source code utilized to generate the results reported in this work are openly available at \footnote{\url{https://github.com/nitrobino/Data_RealSpace_OM}}.

\bibliographystyle{naturemag}
\bibliography{main_text}

\section{End Matter}

\begin{figure*}[t]
\centering \includegraphics[width=1\textwidth]{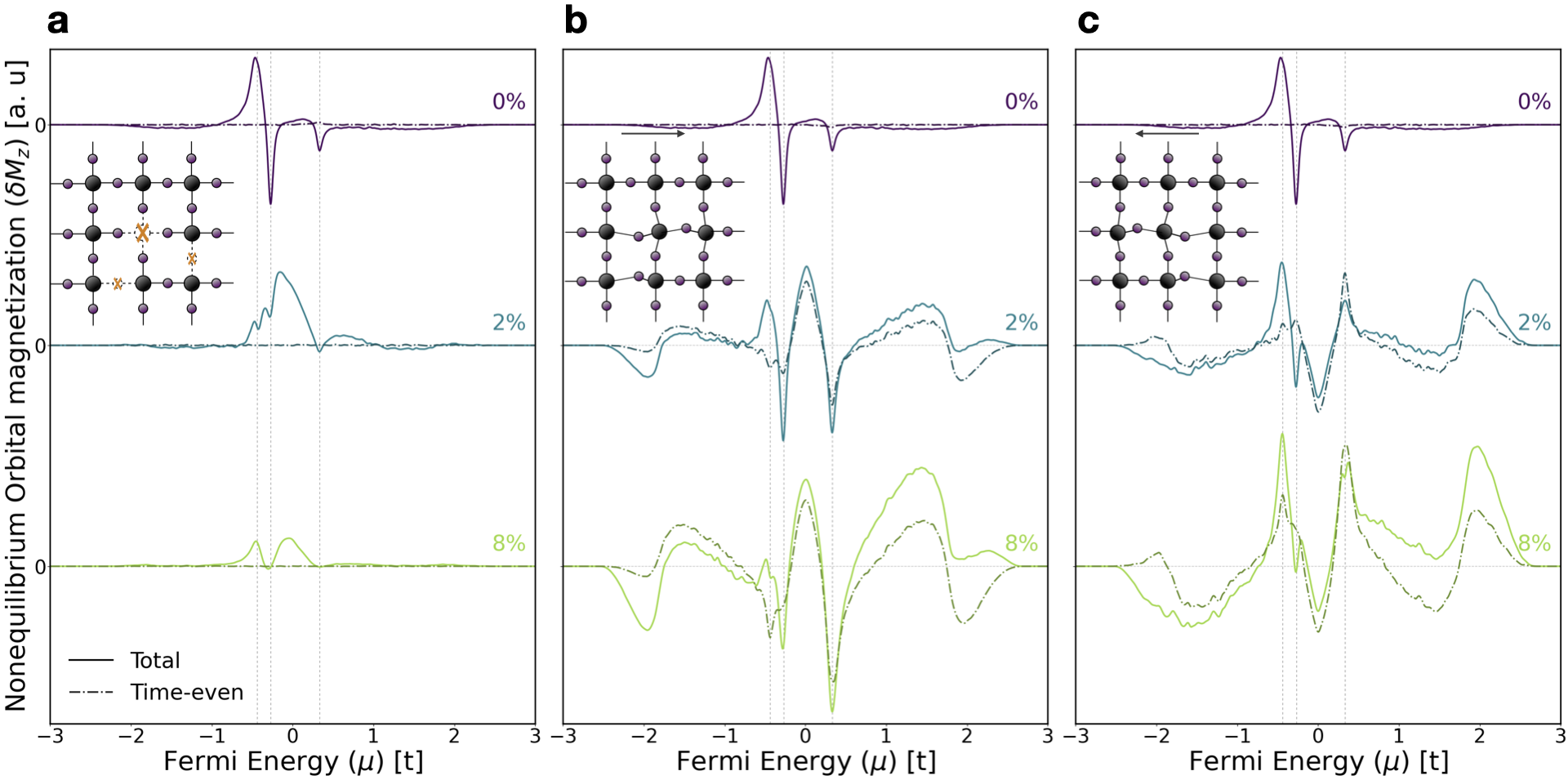} \caption{\textbf{Defect-induced symmetry
breaking effects: total and time-even contributions to $\delta \textbf{M}$}. \textbf{a} Vacancies do not
yield a time-even response, consistent with the
preservation of $\mathcal{PT}$ symmetry on average. \textbf{b} Local
lattice distortions applied to the $p_x$ orbital
generate a finite time-even response even at dilute concentration,
which increases with increasing concentration. \textbf{c} Reversing the
distortion direction produces the symmetry-related response, inverting
the sign of the time-even contribution to $\delta M_z$. Dashed vertical lines
mark the Dirac points of the clean system. Solid lines show the total
response, whilst dash-dotted lines show the time-even contribution.}
\label{Fig3} 
\end{figure*}

\textit{Time-even vs time-odd contributions - }The linear-response magnetization takes the form $\delta M_{b}  = K_{ab}\, E_{a}$, where $K$ is the linear magnetoelectric response tensor. This quantity can be separated into time-odd and time-even components, $K=\tilde\alpha + \beta$. Both contributions follow directly from the symmetric (time-odd) and antisymmetric (time-even) components of $\sigma_{ab,c}$ with respect to the interchange $a\leftrightarrow b$, respectively.  

Within the relaxation time approximation \cite{Salemi_19,Hara_20,He_20,Leiva_23,Burgos_24}, these two sectors capture different microscopic mechanisms of magnetoelectric phenomena: the time-odd  reactive response is associated with the intrinsic mechanism, grounded in the quantum geometry of wavefunctions, whereas the time-even dissipative response (a Fermi-surface property by construction \cite{note_SM}) is governed by extrinsic microscopic mechanisms stemming from disorder and other scattering sources. Such a binary distinction is challenged by our real-space study, which unambiguously demonstrates that the 'intrinsic' (i.e. time-odd) $\delta {\bf M}$ is strongly modified by elastic scattering processes   (see main text). In what follows, we elaborate on the important role played by the spatial symmetries of the disorder ensemble.   

Random distributions of non-magnetic point defects preserve, on average, the $\mathcal P$ (and thus the \(\mathcal{PT}\)) symmetry of the model. Thus, in the thermodynamic (bulk) limit, the time-even response in such cases is inactive. We now show that dilute polar bond disorder activates the otherwise forbidden time-even sector. The results are shown in Fig.~\ref{Fig3} and compared with the vacancy disorder case. As expected from the consideration above, vacancy disorder does not produce a time-even magnetoelectric response, independently of concentration (Fig. \ref{Fig3}a). Conversely, polar disorder generates a sizeable time-even contribution to $\delta{\bf M}$ that scales with the defect concentration (Fig. \ref{Fig3}b). Moreover, inverting the polar distortion reverses the sign of the time-even response (Fig.~\ref{Fig3}c).

These trends can be understood from the symmetry of the disorder
ensemble. A vacancy removes an orbital from the lattice and strongly distorts the local current texture; however, the disorder-averaged response retains $\mathcal{PT}$ symmetry and does not activate the time-even channel (Fig. \ref{Fig3}a). The crucial distinction with the vacancy case is that, for polar defect distribution, a $\mathcal{PT}$ transformation maps a given polar configuration onto one
with the opposite distortion direction, which does not belong to the same disorder ensemble. The ensemble itself therefore breaks
$\mathcal{PT}$ symmetry, removing the cancellation of the time-even sector and  allowing a pure (time-even) Fermi-surface contribution to emerge. The sign reversal observed upon inverting the polar disorder is the corresponding mirror-symmetric counterpart, and an equal mixture of the two polar orientations would restore $\mathcal{PT}$ symmetry and suppress the time-even response.

\textit{Methods - }To overcome the scaling limitations of exact diagonalization, we evaluate
the real-space response functions using the kernel polynomial method as
implemented in the open-source KITE package~\citep{joao_kite_2020}. The relevant spectral operators for the non-equilibrium response, $\Delta \hat G\left(\varepsilon\right)$ and $\partial_{\varepsilon}\hat G^{\rm R}(\varepsilon)$ (generically denoted as $g\left({\cal H}, \varepsilon\right)$) are expanded in a truncated Chebyshev series:
\begin{equation}
    g\left({\cal H},\varepsilon\right)\,=\,N\left(\Delta\right)\,\sum^{M-1}_{n=0}T_{n}\left(\tilde{{\cal H}}\right)c_{n}\left(\tilde{\varepsilon}\right),
\end{equation}
where $T_{n}$ is the Chebyshev polynomial of order  $n$ . The Hamiltonian and
energy are rescaled as
$\tilde{\cal H}={\cal H}/\Delta$ and
$\tilde{\varepsilon}=\varepsilon/\Delta$, with $\Delta$ chosen such that
the spectrum of $\tilde{\cal H}$ lies within the interval $(-1,1)$. The
normalisation factor $N(\Delta)$ and the coefficients
$c_n(\tilde{\varepsilon})$ are fixed by the spectral function being expanded. The lattice Green's functions in Eqs. \eqref{eq:sigma_trace}-\eqref{eq:O_average} are
regularised by means of a finite broadening $\eta\rightarrow0^+$, which sets the
spectral resolution \cite{Ferreira_15}. In this work, the Green's functions were evaluated with $M=1024$ polynomials and $\eta=0.025$.

The macroscopic response functions in Eqs.~\eqref{eq:sigma_trace} and~\eqref{eq:O_average} require the evaluation of traces over the full single-particle Hilbert space. For efficiency, these traces are evaluated using
stochastic trace estimation. We introduce random vectors
$|\xi\rangle=\sum_{i=1}^{N_{\mathrm H}}\xi_i |i\rangle$,
where $N_{\mathrm H}$ is the Hilbert-space dimension and the random
variables $\xi_i$ are uncorrelated, have zero mean and unit variance.
The trace of an operator is then estimated from averages of
$\langle\xi|\hat O|\xi\rangle$ \cite{weise_kernel_2006}.  This procedure reduces the computational
cost of macroscopic observables to linear scaling in $N_{\mathrm H}$ and
benefits from self-averaging in the large systems considered here. All
macroscopic response coefficients are reported as traces per physical
volume, as in Eq.~\eqref{eq:O_average}, rather than per Hilbert-space
dimension.  For the disorder-averaged macroscopic magnetization,
we performed averages of up to $1200$ independent disorder configurations, which was
sufficient to suppress sample-to-sample fluctuations owing to the
self-averaging behavior of $\delta M_z$. The local marker representation in Figures \ref{Fig1}d and \ref{Fig2}b were calculated on regions of up to $128\times128$ unit-cells. The computation of local markers
introduces an additional scaling with the size of the real-space region,
but remains tractable for the system sizes used in this work.

The real-space linear response formalism introduced in this work was applied to a
tight-binding model of a copper-oxide plane on a Lieb lattice
\citep{shinadaQuantumTheoryIntrinsic2023}.  The system
comprises Cu $d$ and O $p_x$, $p_y$
orbitals (Fig.$\,$\ref{Fig1}.a). Following Ref. \citep{shinadaQuantumTheoryIntrinsic2023},
the tight-binding parameters were set to $t = 1.0$, $r = 1.5t$,
$t' = 0.5t$ throughout the simulations.
The impact of translational-symmetry breaking on non-equilibrium orbital
magnetization was studied using two disorder models: vacancy disorder and
polar bond disorder. Vacancy disorder was implemented by removing orbitals
from randomly selected lattice sites. Unless otherwise specified,
vacancies were distributed randomly over the Cu and O sublattices. Polar
bond disorder was introduced by adding an imbalance of magnitude $0.5t$
between the left- and right-hopping amplitudes associated with the
$p_x$ orbital. The point-defect concentration is defined with respect to
the number of unit cells in the corresponding pristine sample.

\appendix

\onecolumngrid

\section*{Supplemental Material}

\section{Appendix A: Local Magnetisation Marker}

\begin{figure}[h]
\includegraphics[width=0.8\textwidth]{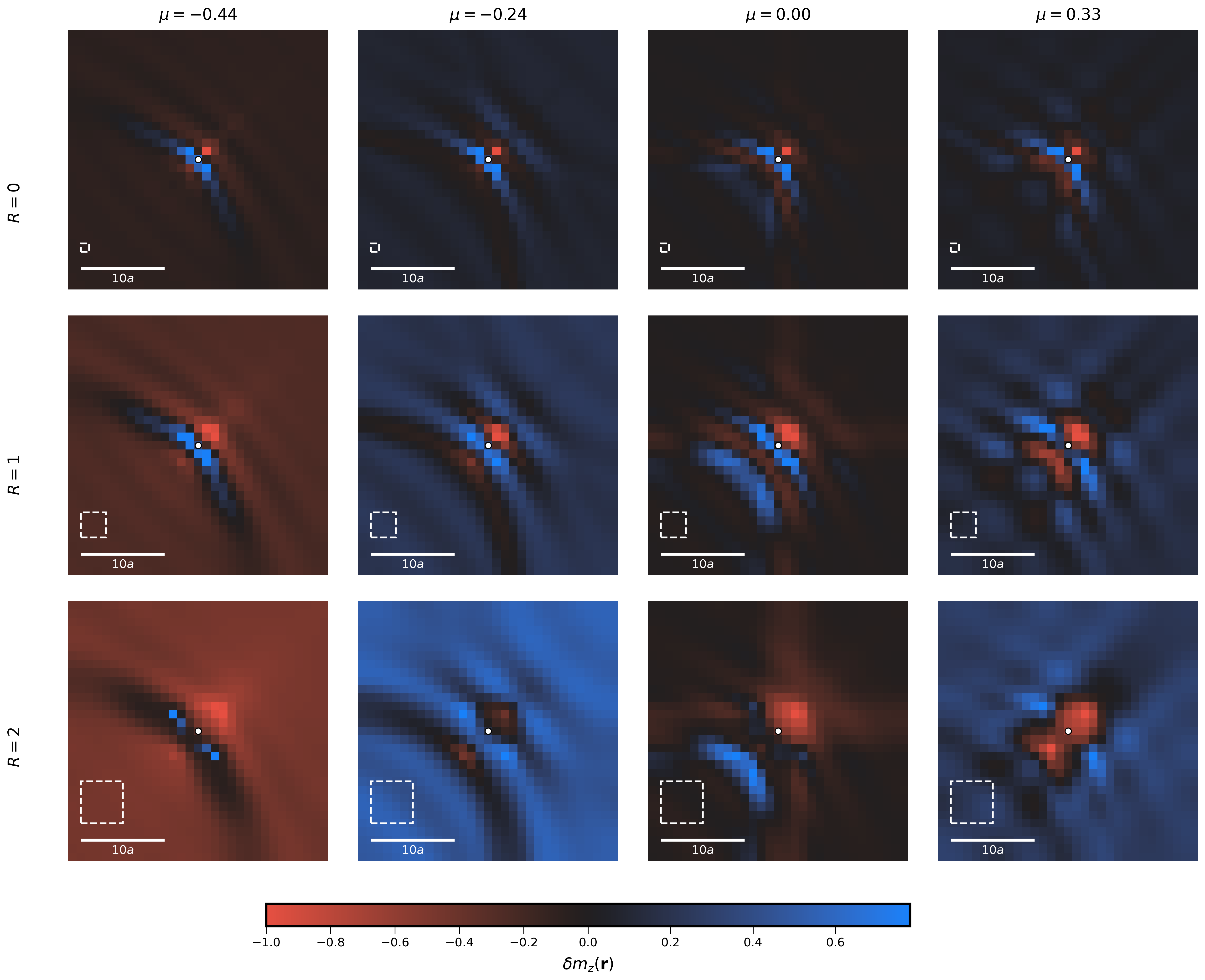}\caption{Spatial maps of the defect-induced local orbital-magnetisation marker
$\delta m_{z}(\mathbf{r})$ for a single vacancy defect, shown for selected values of
chemical potentials $\mu$ and coarse-graining radii $R$. The top
row shows the unsmoothed marker ($R=0$), where microscopic oscillations
associated with the lattice-scale response are clearly visible. Increasing
the averaging radius suppresses these short-wavelength oscillations
and reveals the net sign of the vacancy-induced contribution. White
circles indicate the vacancy position, while the dashed white square
indicates the spatial averaging window used for the coarse-grained
marker. The scale bar corresponds to $10a$. Red and blue denote opposite
signs of $\delta m_{z}(\mathbf{r})$.}\label{fig:Spatial-maps-of}

\end{figure}

Following a similiar reasoning of Ref. \cite{bianco_orbital_2013}, we
may define a local \textit{non-equilibrium} magnetisation marker 
\begin{equation}
\delta m_{a}\left(\mathbf{r}\right)=k_{ab}\left(\mathbf{r}\right)E_{b},
\end{equation}
where $k_{ab}\left(\mathbf{r}\right)$ is obtained from local versions
of conductivity $\sigma^{S}_{ab,c}\left(\mathbf{r}_{\alpha}\right)$,
$\sigma^{A}_{ab,c}\left(\mathbf{r}_{\alpha}\right)$ (see accompanying paper \cite{Pinho_PRR}) and
\begin{equation}
\sigma_{ab,c}\left(\mathbf{r}_{\alpha}\right)=\frac{ie^{2}}{2\pi\hbar}\int d\varepsilon\,f(\varepsilon)\,\mathrm{Re}\,\bra{\mathbf{r}_{\alpha}}\Delta G(\varepsilon)\{R_{c},v_{a}\}\partial_{\varepsilon}G^{R}(\varepsilon)v_{b}-\Delta G(\varepsilon)v_{a}\partial_{\varepsilon}G^{R}(\varepsilon)\{R_{c},v_{b}\}\ket{\mathbf{r}_{\alpha}}.
\end{equation}
Moreover, $\ket{\mathbf{r}_{\alpha}}$ make up the real-space basis. By
virtue of the position operator being diagonal in this basis, we can
still recast this in a commutator form
\begin{equation}
\begin{alignedat}{1}\sigma_{ab,c}\left(\mathbf{r}_{\alpha}\right) & =\frac{ie^{2}}{2\pi\hbar}\int d\varepsilon\,f(\varepsilon)\,\mathrm{Re}\,\bra{\mathbf{r}_{\alpha}}\Delta G(\varepsilon)\left(R_{c}v_{a}+v_{a}R_{c}\right)\partial_{\varepsilon}G^{R}(\varepsilon)v_{b}-\Delta G(\varepsilon)v_{a}\partial_{\varepsilon}G^{R}(\varepsilon)\left(R_{c}v_{b}+v_{b}R_{c}\right)\ket{\mathbf{r}_{\alpha}}\\
 & =\frac{ie^{2}}{2\pi\hbar}\int d\varepsilon\,f(\varepsilon)\,\mathrm{Re}\,\bra{\mathbf{r}_{\alpha}}\Delta G(\varepsilon)\left(R_{c}v_{a}+v_{a}R_{c}\right)\partial_{\varepsilon}G^{R}(\varepsilon)v_{b}-\Delta G(\varepsilon)v_{a}\partial_{\varepsilon}G^{R}(\varepsilon)R_{c}v_{b}-R_{c}\Delta G(\varepsilon)v_{a}\partial_{\varepsilon}G^{R}(\varepsilon)v_{b}\ket{\mathbf{r}_{\alpha}}\\
 & =\frac{ie^{2}}{2\pi\hbar}\int d\varepsilon\,f(\varepsilon)\,\mathrm{Re}\,\bra{\mathbf{r}_{\alpha}}\left[\Delta G(\varepsilon),R_{c}\right]v_{a}\partial_{\varepsilon}G^{R}(\varepsilon)v_{b}+\Delta G(\varepsilon)v_{a}\left[R_{c},\partial_{\varepsilon}G^{R}(\varepsilon)\right]v_{b}\ket{\mathbf{r}_{\alpha}},
\end{alignedat}
\end{equation}
Where we have used $\bra{\mathbf{r}_{\alpha}}\Delta G(\varepsilon)v_{a}\partial_{\varepsilon}G^{R}(\varepsilon)v_{b}R_{c}\ket{\mathbf{r}_{\alpha}}=r_{c}\bra{\mathbf{\mathbf{r}_{\alpha}}}\Delta G(\varepsilon)v_{a}\partial_{\varepsilon}G^{R}(\varepsilon)v_{b}\ket{\mathbf{r}_{\alpha}}=\bra{\mathbf{r}_{\alpha}}R_{c}\Delta G(\varepsilon)v_{a}\partial_{\varepsilon}G^{R}(\varepsilon)v_{b}\ket{\mathbf{r}_{\alpha}}$
with $R_{c}\ket{\mathbf{r}_{\alpha}}=r_{c}\ket{\mathbf{r}_{\alpha}}$
in the last term of step 2. 

\section{Appendix B: Coarse-graining of Local Magnetisation Marker and Vacancy Quantum Interference}

\begin{figure}[h]
\includegraphics[width=0.45\textwidth]{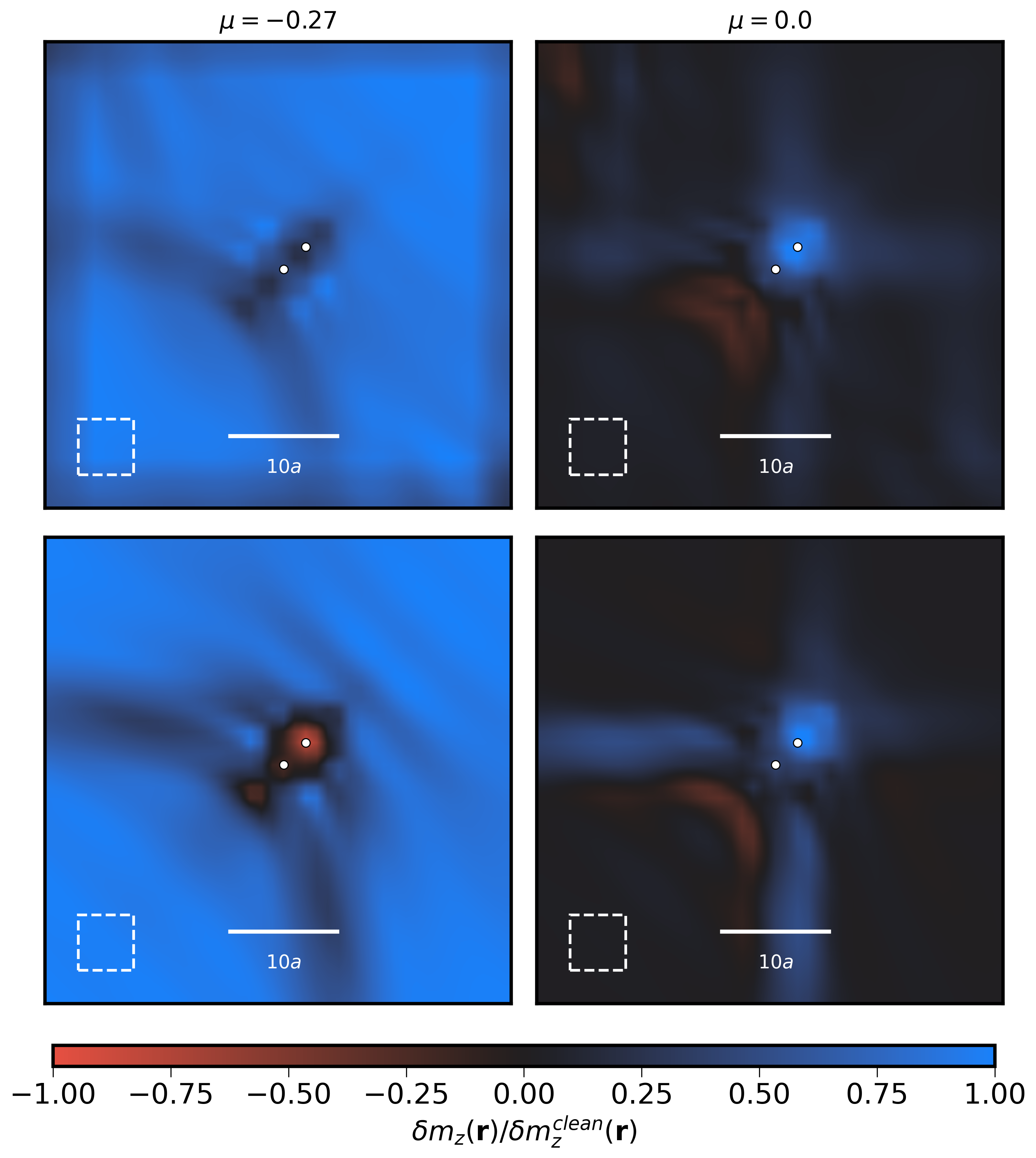}\caption{Spatial maps of the defect-induced local orbital-magnetisation marker
$\delta m_{z}(\mathbf{r})$ for a single vacancy, shown for different
chemical potentials $\mu$ and coarse-graining radii $R$. The top
row shows the unsmoothed marker ($R=0$), where microscopic oscillations
associated with the lattice-scale response are clearly visible. Increasing
the averaging radius suppresses these short-wavelength oscillations
and reveals the net sign of the vacancy-induced contribution. White
circles indicate the vacancy position, while the dashed white square
indicates the spatial averaging window used for the coarse-grained
marker. The scale bar corresponds to $10a$. Red and blue denote opposite
signs of $\delta m_{z}(\mathbf{r})$.}\label{fig:sum_closer_comparison}
\end{figure}

In this section, we report the local magnetisation marker around a $d$
(Cu) vacancy for different chemical potentials and different spatial averaging
windows (Figure \eqref{fig:Spatial-maps-of}). Each pixel corresponds
to the total sum of the three orbitals' contribution in the that unit
cell. Increasing the averaging radius suppresses
these short-wavelength oscillations and reveals the net sign of the vacancy-induced contribution. 

In Figure \ref{fig:sum_closer_comparison}, we also compare the map of two vacancies put close together to the superposition of the response of two isolated vacancies to see if there is occuring interference. We see a clear difference at $\mu=-0.27$, where the sign of the total response was seen to flip for sufficient defect concentration. 

\section{Appendix C: The Impact of Vacancy Type and Placement}

\begin{figure}[!h]
\includegraphics[width=0.9\textwidth]{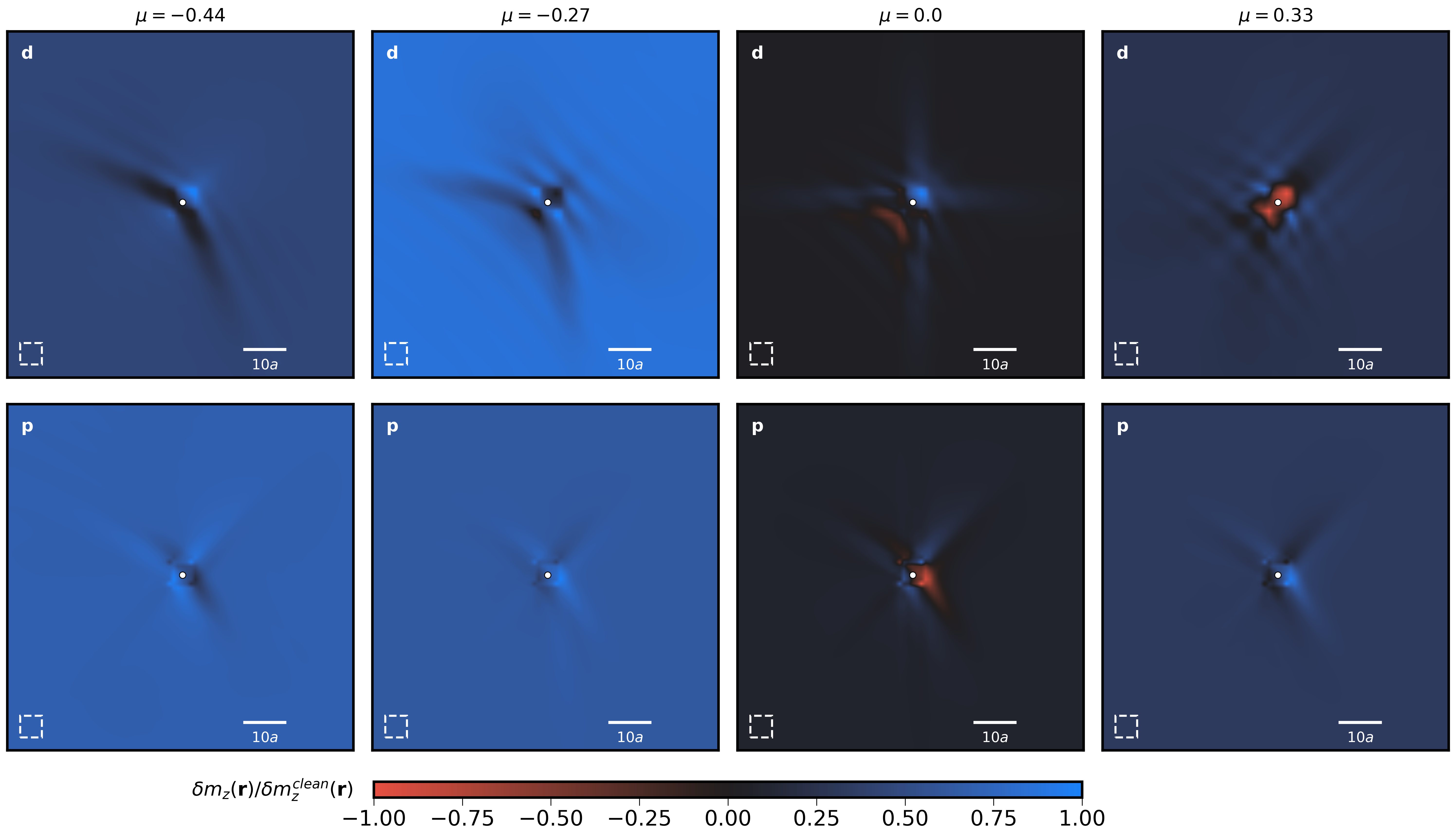}\caption{Spatial maps of the defect-induced local orbital-magnetisation marker
$\delta m_{z}(\mathbf{r})$ for two types of vacancies, shown for different
chemical potentials $\mu$ and coarse-graining radii $R$. The top
row shows the map generated by a single $d$ vacancy, whereas the bottom row shows that of a $p$ vacancy. White
circles indicate the vacancy position, while the dashed white square
indicates the spatial averaging window used for the coarse-grained
marker. The scale bar corresponds to $10a$.}\label{fig:sup_d_vs_p}
\end{figure}

\begin{figure}[h]
\includegraphics[width=0.95\textwidth]{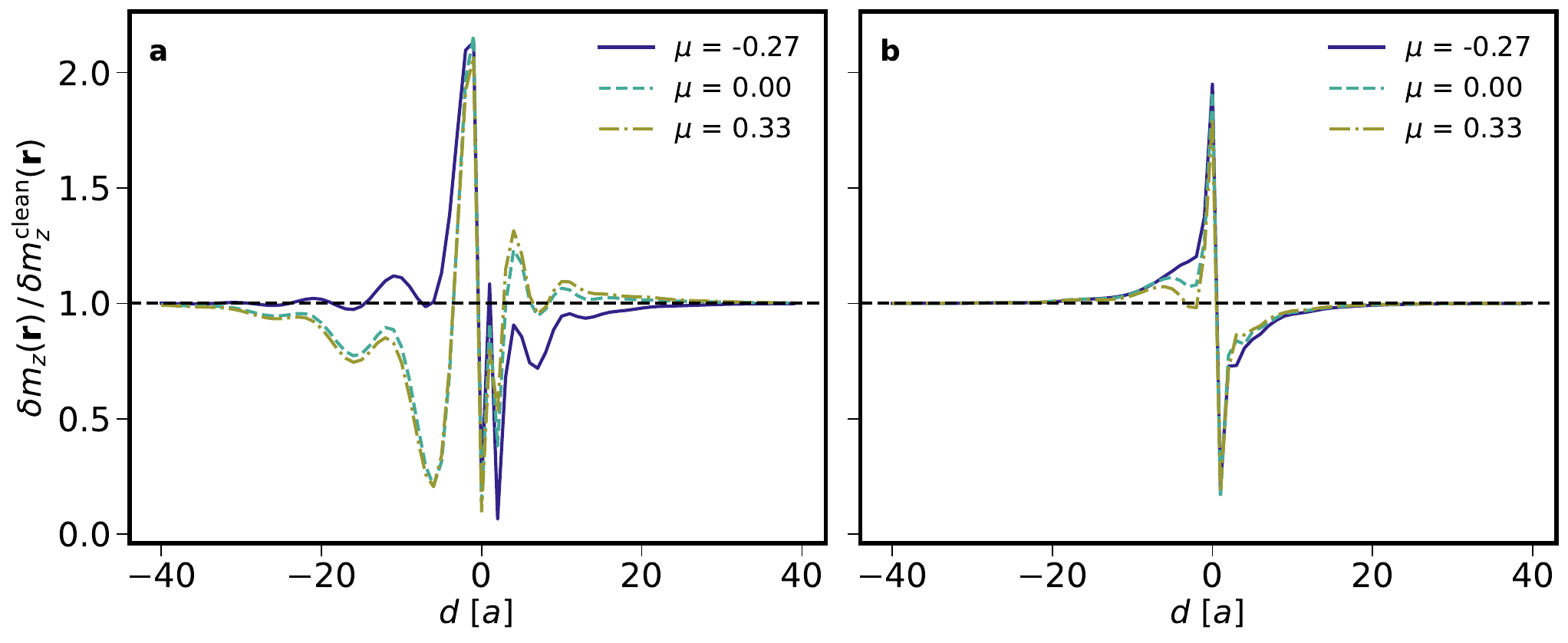}\caption{Real-space profile of the normalized local magnetisation marker, $\delta m_z(\mathbf r)\,/\,\delta m^{\mathrm{clean}}_z (\mathbf r)$, as a function of the distance to a point defect along the longitudinal cross-section. $\,\mathbf a$. $d$ vacancy. $\mathbf b$ $p$ vacancy. For both cases the local marker tends to the intrinsic (pristine) non-equilibrium orbital magnetisation value away from the vacancy.}\label{fig:cutt_decay}
\end{figure}

In this section, we report spatial maps for two types of point defects (Fig. \ref{fig:sup_d_vs_p}) and confirm in both cases that away from the vacancy location, the response tends to that of the bulk crystalline system, as expected (Fig. \ref{fig:cutt_decay}). 

\end{document}